\documentclass[12pt]{article}
\usepackage{amsmath,amsfonts,amssymb}

\usepackage{indentfirst}
\makeatletter
\renewcommand{\@begintheorem}[2]{\begin{trivlist}
\item[\hspace{\labelsep}{\bfseries#1\ #2.}]\itshape}
\renewcommand{\@opargbegintheorem}[3]{\begin{trivlist}
\item[\hspace{\labelsep}{\bfseries#1\ #2\ (#3).}]\itshape}
\renewcommand{\@endtheorem}{\end{trivlist}}
\makeatother
\author{Sergey V.\,Smirnov\thanks{Department of Mathematics and Mechanics, Moscow State University. E-mail: {\tt ssmirnov@higeom.math.msu.su}}}
\title{Integral preserving discretization of 2D Toda lattices}

\def\pa{\partial}

\def\phi{\varphi}

\newtheorem{proposition}{\sc Proposition}
\newtheorem{theorem}{\sc Theorem}
\newtheorem{remark}{\sc Remark}
\newtheorem{example}{\sc Example}

\newtheorem{definition}{\sc Definition}
\newtheorem{lemma}{\sc Lemma}

\begin{document}

\maketitle
\begin{abstract}
There are different methods of discretizing integrable systems. We consider semi-discrete analog of two-dimensional Toda lattices associated to the Cartan matrices
of simple Lie algebras that was proposed by Habibullin in 2011. This discretization is based on the notion of Darboux integrability. Generalized Toda lattices are
known to be Darboux integrable in the continuous case (that is, they admit complete families of characteristic integrals in both directions). We prove that
semi-discrete analogs of Toda lattices associated to the Cartan matrices of all simple Lie algebras are Darboux integrable. By examining the properties of
Habibullin's discretization we show that if a function is a characteristic integral for a generalized Toda lattice in the continuous case, then the same function
is a characteristic integral in the semi-discrete case as well. We consider characteristic algebras of such integral-preserving discretizations of Toda lattices
to prove the existence of complete families of characteristic integrals in the second direction.
\end{abstract}

\section{Introduction}

Two-dimensional Toda lattice
\begin{equation}
\label{qt}
q_{i,xy}=\exp(q_i-q_{i+1})-\exp(q_{i-1}-q_i)
\end{equation}
plays an important role both in classical differential geometry and in mathematical physics. It is known to be Darboux integrable (i.e. it admits complete family
of essentially independent characteristic integrals), if the trivial boundary conditions $q_{-1}=-\infty$ and $q_{r+1}=+\infty$ are imposed for some natural $r$.
Toda lattice can be rewritten in the form
\begin{eqnarray}
\left\{
\begin{array}{l}
u^1_{xy}=\exp(2u^1-u^2)\\
u^i_{xy}=\exp(-u^{i-1}+2u^i-u^{i+1}),\quad i=2,\dots,r-1\\
u^r_{xy}=\exp(-u^{r-1}+2u^r)
\end{array}
\right.,
\end{eqnarray}
where $q_i=u^{i+1}-u^i$. This system is a particular case of the so-called {\it exponential systems}
\begin{equation}
\label{expsyst}
u^i_{xy}=\exp\left(\sum\limits_{j=1}^r a_{ij} u^j\right),\quad i=1,2,\dots, r,
\end{equation}
that were introduced in~\cite{ShJa} (here $a_{ij}$ are constant coefficients). Such integrable generalizations of the Toda system corresponding to the Cartan matrices
$M=(a_{ij})$ of all simple Lie algebras were studied in a number of papers in the beginning of 1980-ies~\cite{ShJa}-\cite{LSSh}. Almost at the same time, discrete
versions of some of these systems started to appear in literature within the frame of discretizing the theory of integrable systems~\cite{Hi1,Hi2}. Due to the fact
that generalized Toda systems are related to many relevant mathematical theories, the number of papers on the subject is enormous: continuous Toda
systems and their discretizations are addressed from different perspectives and studied using various methods. Not even attempting to make a review, we will mention
the papers~\cite{Wa}-\cite{YF} that are important for our approach to the problem of discretizeing Toda lattices.

Purely discrete versions of the generalized Toda lattice, associated with the $A$-series Cartan matrices, were studied in~\cite{Wa,Do,AS}. Purely discrete and
semi-discrete versions of the $C$-series Toda lattice were examined in~\cite{Ha06,Sm12}. Discretizations of generalized Toda lattices introduced in these papers
are based mostly on the notion of the Laplace invariants of hyperbolic difference or differential-difference operators and on their specific properties, and therefore
this approach is not applicable in the general case of an exponential system associated to arbitrary Cartan matrix.

There are several general approaches that lead to (semi)-discrete versions of exponential systems~(\ref{expsyst}) corresponding to Cartan matrices.
Hamiltonian approach was used in~\cite{IH} to obtain semi-discrete analogs of exponential systems associated to Cartan matrices of all simple Lie algebras.
Another approach to the problem of discretization of exponential systems was proposed in~\cite{HZY,GHY}. The idea is to look for a semi-discrete
system for the functions $u^i_n$, depending on continuous variable $x$ and discrete variable $n$, such that its characteristic $n$-integrals are given by the same
formulas as $y$-integrals of the continuous model. We will call this method {\it integral preserving discretization}. This approach had appeared to be fruitful
earlier in the study of Darboux integrable semi-discrete scalar equations~\cite{HZhS}, before it was applied to the Toda lattices. The analysis of the case $r=2$
from the viewpoint of integral preserving method allowed the authors~\cite{HZY} to propose the following discretization for exponential system~(\ref{expsyst})
with the coefficient matrix $M=(a_{ij})$:
\begin{equation}
\label{sdexpsyst}
u^i_{n+1,x}-u^i_{n,x}=\exp\left(\sum\limits_{j=1}^{i-1} a_{ij} u^j_n+\frac{a_{ii}}{2}(u^i_n+u^i_{n+1})+
\sum\limits_{j=i+1}^{r} a_{ij} u^j_{n+1}\right).
\end{equation}
Similar approach~\cite{GHY} allows to define the analogs of exponential systems in the purely discrete case:
\begin{multline}
\label{pdexpsyst}
\exp(u^i_{n+1,m+1}+u^i_{n,m}-u^i_{n+1,m}-u^i_{n,m+1})=\\
=1+\exp\left(\sum\limits_{j=1}^{i-1} a_{ij} u^j_{n,m+1}+\frac{a_{ii}}{2}(u^j_{n,m+1}+u^j_{n+1,m})+
\sum\limits_{j=i+1}^{r} a_{ij} u^j_{n+1,m}\right).
\end{multline}
In both cases, Darboux integrability was proved only for discretized systems associated with all Cartan matrices of the rank $2$~\cite{HZY,GHY}. In the general
case, Darboux integrability for (semi)-discrete versions of the Toda lattice is proven only for the $A$- and $C$-series lattice by constructing a generating
function for characteristic integrals~\cite{Sm15}, but this method is not applicable for the $B$-, $D$-series lattices and for the systems corresponding to
exceptional Cartan matrices $E_6$--$E_8$ and $F_4$.

Although discterizations of $A$-series Toda lattice given in~\cite{IH} and in~\cite{HZY} are the same, the methods used there produce different semi-discrete systems
for other series of Cartan matrices (continuum limits in both cases are the same). Another version of semi-discrete Toda system of the series $B$ is obtained
in~\cite{CHHL} by considering a modification of skew-orthogonal polynomials that arise in the random matrix theory. In~\cite{YF} direct linearization method was used
to study semi-discrete analogs of Toda systems corresponding to some series of affine Cartan matrices. In the purely discrete case there also exist different
versions of Toda systems corresponding to Cartan matrices, see~\cite{GHY,KNS}.

Various properties of integrable models are usually taken as a basis for finding discretizations. Discretization of Toda systems proposed in~\cite{HZY} is based on
the notion of Darboux integrability and hence it is important to show that systems introduced in~\cite{HZY} are Darboux integrable indeed. In this paper we focus
on semi-discrete exponential systems~(\ref{sdexpsyst}) and prove that such systems corresponding to the Cartan matrices of all simple Lie
algebras are Darboux integrable. Therefore we justify the integral preserving discretization method for generalized Toda lattices. More precisely, we prove that if
exponential system~(\ref{expsyst}) admits $y$-integral
$$
I=I(u^1_x,\dots,u^r_x,u^1_{xx},\dots,u^r_{xx},u^1_{xxx},\dots,u^r_{xxx},\dots),
$$
then the same function
$$
I_n=I(u^1_{n,x},\dots,u^r_{n,x},u^1_{n,xx},\dots,u^r_{n,xx},u^1_{n,xxx},\dots,u^r_{n,xxx},\dots),
$$
whose arguments are replaced by the dynamical variables for the semi-discrete case,
is an $n$-integral for discretization~(\ref{sdexpsyst}) of this system. Besides this, using {\it characteristic algebras}, we prove that semi-discrete
versions~(\ref{sdexpsyst}) of $B$-, $D$-series Toda lattices and the systems associated to exceptional Cartan matrices $E_6$--$E_8$ and $F_4$ admit complete families
of essentially independent characteristic $x$-integrals (Darboux integrability of $A$- and $C$-series lattices have been proved earlier). Altogether, this proves
Darboux integrability of all semi-discrete Toda lattices~(\ref{sdexpsyst}) corresponding to the Cartan matrices of simple Lie algebras, which was conjectured
in~\cite{HZY}.

The paper is structured as follows: in Section~\ref{darbint} we review the notions of Darboux integrability, characteristic algebra and the relation between them.
In Section~\ref{intpres} we describe Habibullin's method and prove the existence of a complete family of characteristic $n$-integrals for semi-discrete
lattices~(\ref{sdexpsyst}) corresponding to the Cartan matrices of all simple Lie algebras and hence we justify Habibullin's integral preserving discretization
method by showing that if a function is a $y$-integral of discrete exponential system, then the same function defines an $n$-integral for its semi-discrete analog.
Basic properties of characteristic algebras for exponential systems associated to the Cartan matrices of simple Lie algebras are discussed in Section~\ref{charcont}.
In Section~\ref{xint} we prove the existence of a complete family of independent $x$-integrals for semi-discrete exponential systems corresponding to the Cartan
matrices of all simple Lie algebras.

\section{Darboux integrability and characteristic algebras}\label{darbint}

In the theory of integrable systems there are several different approaches to integrability depending in the class in the systems that are being considered:
Liouville integrability, existence of a Lax pair, existence of higher symmetries. {\it Darboux integrability} is a kind of ``very strong'' integrability
that is defined for hyperbolic systems and that is closely related to explicit integrability. We start this Section with a series of definitions and
notation~\cite{ShJa,SS} that will be used in this paper.

Function
$$
I=I(u^1,\dots,u^r, u^1_x,\dots,u^r_x,u^1_{xx},\dots,u^r_{xx},u^1_{xxx},\dots,u^r_{xxx},\dots)
$$
is called a {\it $y$-integral} of hyperbolic system
\begin{equation}
\label{hypeqn}
u^i_{xy}=F^i(x,y,u^1,\dots, u^r,u^1_x,\dots,u^r_x,u^1_y,\dots,u^r_y),\quad i=1,\dots,r,
\end{equation}
if its total derivative with respect to $y$ by virtue of the system vanishes:
$$
0=D_y(I)=\sum\limits_{i=1}^r \left(u^i_y\frac{\pa I}{\pa u^i}+F^i\frac{\pa I}{\pa u^i_x}+D_x(F^i)\frac{\pa I}{\pa u^i_{xx}}+
D^2_x(F^i)\frac{\pa I}{\pa u^i_{xxx}}+\dots\right),
$$
where $D_x$ is the total derivative with respect to $x$. The highest order of $x$-derivative of the functions $u^1,\dots u^r$, on which $y$-integral $I$ depends,
is called {\it the order} of $I$; $y$-integral is called {\it non-trivial} if it depends not only on the independent variable $x$. Here and further we will
consider only non-trivial integrals; {\it $x$-integrals} of hyperbolic system~(\ref{hypeqn}) are defined similarly. Both $x$- and $y$-integrals are
called {\it characteristic integrals}. Denote
$$
u^i_1=u^i_x,\quad u^i_2=u^i_{xx},\quad u^i_3=u^i_{xxx},\dots,\quad i=1,\dots,r.
$$
Characteristic $y$-integrals $I_1,\dots, I_k$ of orders $d_1,\dots d_k$ are called {\it essentially independent} if the rank of the matrix
\begin{eqnarray}
\nonumber
\left(
\begin{array}{cccc}
\frac{\pa I_1}{\pa u^1_{d_1}} & \frac{\pa I_1}{\pa u^2_{d_1}} & \dots & \frac{\pa I_1}{\pa u^r_{d_1}}\\
\frac{\pa I_2}{\pa u^1_{d_2}} & \frac{\pa I_2}{\pa u^2_{d_2}} & \dots & \frac{\pa I_2}{\pa u^r_{d_2}}\\
\vdots & &\ddots &\\
\frac{\pa I_k}{\pa u^1_{d_k}} & \frac{\pa I_k}{\pa u^2_{d_k}} & \dots & \frac{\pa I_k}{\pa u^r_{d_k}}\\
\end{array}
\right)
\end{eqnarray}
is equal to $k$.
\begin{definition}
\rm
Hyperbolic system~(\ref{hypeqn}) is called {\it Darboux integrable} if it admits complete families of essentially independent $x$- and $y$-integrals.
\end{definition}
\begin{example}
\rm
Liouville equation
\begin{equation}
\label{eqliouv}
u_{xy}=\exp u
\end{equation}
is Darboux integrable since it admits characteristic integrals in both directions: functions
\begin{equation}
\label{intli}
I=u_{xx}-\frac{1}{2}u_x^2\quad\hbox{and}\quad J=u_{yy}-\frac{1}{2}u_y^2
\end{equation}
are $y$- and $x$-integrals respectively.
\end{example}

Exponential systems associated with the Cartan matrices of simple Lie algebras (i.e. exponential systems~(\ref{expsyst}) such that the coefficient matrix is the
Cartan matrix of one of simple Lie algebras) are known to be Darboux integrable~\cite{Le}. These systems are also known in literature as generalized Toda lattices
corresponding to the Cartan matrices of simple Lie algebras. Explicit formulas for characteristic integrals in terms of wronskians were found in~\cite{De} for
Toda lattices of series $A$--$D$. Another approach that allows to obtain generating function for characteristic integrals was developed in~\cite{Ni} for
the $A$-series Toda lattices and in~\cite{Sm15} for lattices of the series $A$--$C$.

Characteristic integrals are two-dimensional analogs of first integrals for ODEs, but there is an essential difference between these two cases: hyperbolic
equations having characteristic integrals are exceptional. If function $I$ is a $y$-integral of~(\ref{hypeqn}), then its $x$-derivatives $D_x(I),D^2_x(I),\dots$
are obviously also $y$-integrals, but these integrals are not essentially independent. The existence of a complete family of characteristic integrals is controlled
by an algebraic tool --- Lie algebra of differential operators that is called {\it the characteristic algebra} of hyperbolic system~\cite{ZISh,ShJa,LSSh}.
Characteristic algebra can be defined for arbitrary hyperbolic system of form~(\ref{hypeqn}), but in this general case it should be considered as a Lie-Rinehart
algebra (see discussion in~\cite{MS}). In the special case of exponential systems~(\ref{expsyst}) the situation is more simple and the characteristic algebra
can be referred to as a Lie algebra generated by differential operators of a certain kind.
\begin{definition}
\rm
Lie algebra generated by operators
$$
\frac{\pa}{\pa u^1},\dots,\frac{\pa}{\pa u^r},\quad
D_y=\sum\limits_{i=1}^r \left(e^{w^i}\frac{\pa }{\pa u^i_1}+D_x\left(e^{w^i}\right)\frac{\pa }{\pa u^i_2}+D^2_x\left(e^{w^i}\right)\frac{\pa }{\pa u^i_3}+\dots\right),
$$
where $w^i=a_{i1}u^1+\dots+a_{ir} u^r$, is called {\it the characteristic algebra} of exponential system~(\ref{expsyst}).
\end{definition}

One can easily show that $y$-integrals of exponential systems~(\ref{expsyst}) cannot depend on $u^1,\dots u^r$: they depend only on their $x$-derivatives. Therefore
any $y$-integral annihilates the whole characteristic algebra.
\begin{remark}
\rm
In the general case of hyperbolic systems~(\ref{hypeqn}), one has to define characteristic algebra in the direction of the variable $x$ and characteristic algebra
in the direction $y$. For exponential systems~(\ref{expsyst}), these Lie algebras are isomorphic since variables $x$ and $y$ enter the equations symmetrically.
\end{remark}
\begin{proposition}
Let the matrix $M=(a_{ij})$ of exponential system~(\ref{expsyst}) be non-degenerate. Then its characteristic algebra is generated by operators
$$
\frac{\pa}{\pa u^i},\quad\tilde X_i=e^{w^i}\left(\frac{\pa}{\pa u^i_1}+b^i_1\frac{\pa}{\pa u^i_2}+b^i_2\frac{\pa}{\pa u^i_3}+\dots\right),\quad i=1,\dots,r,
$$
where $b^i_k=b^i_k (w^i_1,w^i_2,\dots w^i_k)=e^{-w^i}D^k_x(e^{w^i})$ is the $k$-th complete Bell polynomial of the variables $w^i_1,w^i_2,\dots w^i_k$ and
$w^i_k=D^k_x(w^i)$.
\end{proposition}

{\bf Proof}. Simple calculation shows that for all $i=1,\dots,r$
$$
\left[\frac{\pa}{\pa u^i},D_y\right]=a_{i1}\tilde X_1+a_{i2}\tilde X_2+\dots+a_{ir}\tilde X_r.
$$
Hence, all operators of the form $a_{i1}\tilde X_1+a_{i2}\tilde X_2+\dots+a_{ir}\tilde X_r$ belong to the characteristic algebra, and it follows from non-degeneracy
of the matrix $M$ that operators $\tilde X_i$ are linear combinations of these operators. Therefore, they belong to characteristic algebra, and since
$D_y=\tilde X_1+\dots+\tilde X_r$, they generate the characteristic algebra together with $\frac{\pa}{\pa u^i}$, where $i=1,\dots,r$. $\Box$

\begin{theorem}{\rm\cite{ShJa}}\label{thchar}
Exponential system~(\ref{expsyst}) is Darboux integrable if and only if its characteristic algebra is finite-dimensional.
\end{theorem}
\begin{example}
\rm
Characteristic algebra of the Liouville equation~(\ref{eqliouv}) is two-dimensional: $\left[\frac{\pa}{\pa u},D_y\right]=D_y$.
\end{example}
\begin{remark}
\rm
Since
$$
\left[\frac{\pa}{\pa u^j},\tilde X_i\right]=a_{ij}\tilde X_i
$$
for all $i,j=1,\dots,r$ and characteristic $y$-integrals of an exponential system~(\ref{expsyst}) cannot depend on $u^1,\dots,u^r$, it is sufficient for the study of
Darboux integrability to consider {\it reduced characteristic algebra} generated by $\tilde X_1,\dots,\tilde X_r$: obviously, exponential
system~(\ref{expsyst}) is Darboux integrable if and only if its reduced characteristic algebra is finite-dimensional. Note that this Lie algebra is
isomorphic to Lie algebra generated by vector fields $X_1,\dots X_r$ where $X_i=e^{-w^i}\tilde X_i$ for all $i=1,\dots r$.
\end{remark}

The notion of Darboux integrability can be extended to the case of (semi)-discrete hyperbolic systems. Function
$$
I_n=I(u_n^1,\dots,u_n^r,u^1_{n,x},\dots,u^r_{n,x},u^1_{n,xx},\dots,u^r_{n,xx},u^1_{n,xxx},\dots,u^r_{n,xxx},\dots)
$$
is called an {\it $n$-integral} of semi-discrete hyperbolic system
\begin{equation}
\label{sdhypeqn}
u^i_{n+1,x}-u^i_{n,x}=F^i (x,n,u^1_n,\dots, u^r_n,u^1_{n,x},\dots,u^r_{n,x},u^1_{n+1},\dots,u^r_{n+1}),\quad i=1,\dots,r,
\end{equation}
if its total difference derivative by virtue of the system vanishes: $I_{n+1}-I_n=0$. One can verify that in the semi-discrete case $n$-integrals cannot depend
on shifted variables $u^i_{n+1}$ and $x$-integrals cannot depend on the derivatives $u^i_{n,x}$, where $i=1,\dots,r$. {\it The order} $d$ of an $x$-integral is defined
as the highest shift $u_{n+d}$ on which it depends. Family of $x$-integrals $J_1,\dots J_k$ of orders $d_1,\dots d_k$ are called {\it essentially independent} if the
rank of the matrix
\begin{eqnarray}
\nonumber
\left(
\begin{array}{cccc}
\frac{\pa J_1}{\pa u^1_{n+d_1}} & \frac{\pa J_1}{\pa u^2_{n+d_1}} & \dots & \frac{\pa J_1}{\pa u^r_{n+d_1}}\\
\frac{\pa J_2}{\pa u^1_{n+d_2}} & \frac{\pa J_2}{\pa u^2_{n+d_2}} & \dots & \frac{\pa J_2}{\pa u^r_{n+d_2}}\\
\vdots & &\ddots &\\
\frac{\pa J_k}{\pa u^1_{n+d_k}} & \frac{\pa J_k}{\pa u^2_{n+d_k}} & \dots & \frac{\pa J_k}{\pa u^r_{n+d_k}}\\
\end{array}
\right)
\end{eqnarray}
is equal to $k$. Similarly to the continuous case, hyperbolic system~(\ref{sdhypeqn}) is called {\it Darboux integrable} if it admits complete families of
essentially independent $n$- and $x$- integrals. In the entirely discrete case, Darboux integrability of hyperbolic system
$$
u^i_{n+1,m+1}-u^i_{n+1,m}-u^i_{n,m+1}+u^i_{n,m}=F^i (n,m,u^1_{n,m},\dots, u^r_{n,m},u^1_{n+1,m},\dots,u^r_{n+1,m},u^1_{n,m+1},\dots,u^r_{n,m+1}),
$$
where $i=1,\dots,r$, requires the existence of essentially independent families of $m$- and $n$-integrals.

Darboux integrability of semi-discrete and entirely discrete exponential systems corresponding to the Cartan matrices of the series $A$ and $C$ (i.e. generalized
(semi)-discrete Toda lattices of the series $A$ and $C$) was proved in~\cite{Sm15}. Another approach allowing to obtain a complete family of essentially independent
$x$-integrals for semi-discrete $A$-series Toda lattice in terms of {\it casoratians} was developed in~\cite{DT}.

\section{Integral preserving discretization}\label{intpres}

There are many different ways to discretize integrable systems. One of the popular methods to discretize a PDE with two independent variables is to consider
iterations of its B\"acklund transformations as shinfts in a new discrete variable. Then the formula for B\"acklund transformation that relates the unknown function $u$ with
its B\"acklund-image $u_1$ is a differential-difference equation and it can be referred to as a semi-discrete analog of the initial PDE. In this case, the
superposition formula plays the role of entirely discrete analog. Although this approach is widely used in the theory of integrable systems, we will use another
approach proposed by Habibullin et al.~\cite{HZhS} that is based on the existence of characteristic integrals for generalized Toda lattices, which is a very
specific property for this class of systems.

In this section, we describe the integral preserving discretization method for exponential systems associated to the Cartan matrices of simple Lie algebras and
justify it by proving that this approach provides Darboux integrable semi-discrete systems.

Goursat~\cite{Go1} had found a complete list of scalar hyperbolic equations having both characteristic integrals of order not greater than $2$
(the so-called {\it Goursat list}) within the frame of the study of hyperbolic equations that admit characteristic integrals. The idea proposed in~\cite{HZhS}
is to take characteristic $y$-integral for an equation in the Goursat list and to find semi-discrete hyperbolic equation such that this function is its $n$-integral.
The discretization obtained using this method inherits the main property of its continuous counterpart: by construction, it admits an $n$-integral of order not
greater than $2$. Surprisingly, all semi-discrete equations found in~\cite{HZhS} appear to be Darboux integrable, i.e. in addition to $n$-integrals they also
admit characteristic $x$-integrals. Moreover, this method was also applied to these semi-discrete equations in order to get entirely discrete analogs of the
equations from the Goursat list~\cite{HZhS}.

The same method was used in~\cite{HZY} to guess an appropriate formula for semi-discrete analog of exponential systems~(\ref{expsyst}) corresponding to the Cartan
matrices of simple Lie algebras. More precisely, careful analysis of the systems corresponding to the Cartan matrices of the rank $2$ and their characteristic
integrals allowed the authors to find semi-discrete analogs for exponential systems of the rank $2$, to obtain formula~(\ref{sdexpsyst}) and to conjecture that
systems~(\ref{sdexpsyst}) are Darboux integrable for the Cartan matrices of all simple Lie algebras. The following theorem is the main result of this section and
it proves the first part of the conjecture from~\cite{HZY} (that Habilullin's discretization preserves $y$-integrals); the second part that states the existence
of sufficient number of essentially independent $x$-integrals is proved in section~\ref{xint}.

\begin{theorem}\label{thmmain}
Let $I=I(u^1_x,\dots,u^r_x,u^1_{xx},\dots,u^r_{xx},u^1_{xxx},\dots,u^r_{xxx},\dots)$ be a $y$-integral of exponential system~(\ref{expsyst}) with non-degenerate
matrix $M=(a_{ij})$.
Then the same function
$$
I_n=I(u^1_{n,x},\dots,u^r_{n,x},u^1_{n,xx},\dots,u^r_{n,xx},u^1_{n,xxx},\dots,u^r_{n,xxx},\dots)
$$
is an $n$-integral of discretization~(\ref{sdexpsyst}) of this system.
\end{theorem}

{\bf Proof}.

Let $I$ be a $y$-integral or order $d$ of the continuous system~(\ref{expsyst}). Similarly to the continuous case, denote
$u^i_{n,k}=D_x^k (u^i_n)$ and $w^i_n=a_{i1}u^i_n+\dots+a_{ir}u^i_r$ for all $i=1,\dots,k$. Using Taylor's expansion, rewrite the difference $I_{n+1}-I_n$,
where variables $u^i_n$ satisfy semidiscrete system~(\ref{sdexpsyst}):
\begin{align}
\notag
I_{n+1}-I_n&=\sum_{i=1}^r\sum_{k=1}^d \frac{\pa I}{\pa u^i_k}\left(u^i_{n+1,k}-u^i_{n,k}\right)+\\
\notag
&+\frac{1}{2}\sum_{i_1,i_2=1}^r\sum_{k_1,k_2=1}^d \frac{\pa^2 I}{\pa u^{i_1}_{k_1}\pa u^{i_2}_{k_2}}\left(u^{i_2}_{n+1,k_2}-u^{i_2}_{n,k_2}\right)
\left(u^{i_1}_{n+1,k_1}-u^{i_1}_{n,k_1}\right)+\dots=\\
\label{abb}
&=\sum\limits_{s=1}^\infty\left(\frac{1}{s!}\sum_{\lambda_{ik}:\,\sum\limits_{i=1}^r\sum\limits_{k=1}^d\lambda_{ik}=s}^{}
\frac{s!}{\prod\limits_{i=1}^r\prod\limits_{k=1}^d\lambda_{ik}!}\prod\limits_{i=1}^r\prod\limits_{k=1}^d\left(u^i_{n+1,k}-u^i_{n,k}\right)^{\lambda_{ik}}
\frac{\pa^s I}{(\pa u^1_1)^{\lambda_{11}}\dots(\pa u^r_d)^{\lambda_{rd}}}\right)
\end{align}
Denote
$$
E_i=\exp(w^i_n),\quad\Delta_i=\exp\left(\frac{a_{ii}}{2}\left(u^i_{n+1}-u^i_n\right)+a_{i,i+1}\left(u^{i+1}_{n+1}-u^{i+1}_n\right)+\dots+
a_{ir}\left(u^r_{n+1}-u^r_n\right)\right),
$$
$t_i=E_i\Delta_i,$ where $i=1,\dots,r$ (here we omit the dependence on $n$ for simplicity in notation). Hence, the $i$-th equation of~(\ref{sdexpsyst})
can be rewritten as $u^i_{n+1,x}-u^i_{n,x}=t_i$ and therefore we need to
simplify the terms of the form $D_x^{k-1}(t_i)$ in~(\ref{abb}) using equations~(\ref{sdexpsyst}).

The remaining part of the proof of this Theorem is divided into a series of propositions on some specific properties of the the functions $t_i$ and the operators
$$
X_i=e^{-w^i}\tilde X_i=\frac{\pa}{\pa u^i_1}+b^i_1\frac{\pa}{\pa u^i_2}+b^i_2\frac{\pa}{\pa u^i_3}+\dots
$$
\begin{proposition}
For all $i=1,\dots,r$ functions $t_i$ satisfy the relation
\begin{equation}
\label{add}
D_x(t_i)=t_i\left(b^i_1+\frac{a_{ii}}{2}t_i+a_{i,i+1}t_{i+1}+\dots+a_{ir}t_r\right).
\end{equation}
\end{proposition}
{\bf Proof}. Differentiate $t_i$ and use relations
$$
D_x(E_i)=b^i_1 E_i,\quad D_x(\Delta_i)=\frac{a_{ii}}{2}t_i+a_{i,i+1}t_{i+1}+\dots+a_{ir}t_r.\quad\Box
$$
\begin{proposition}\label{proprel}
Operators $X_j$ satisfy the relation
$$
X_j (D_x+b^i_1)=(D_x+b^i_1+b^j_1)X_j+a_{ij}
$$
for all $i,j=1,\dots,r$.
\end{proposition}

{\bf Proof}. Complete Bell polynomials $B_k=B_k(v_1,\dots,v_k)$ are known to satisfy the relation
$$
D_x(B_k)=B_{k+1}-B_1 B_k,\quad k=1,2,\dots
$$
Apply this relation to calculate the commutator:
\begin{align*}
\left[X_j,D_x+b^i_1\right]&=\left(\frac{\pa}{\pa u^j_1}+b^j_1\frac{\pa}{\pa u^j_2}+b^j_2\frac{\pa}{\pa u^j_3}+\dots\right)
\left(b^i_1+\sum_{l=1}^r\left(u^l_2\frac{\pa}{\pa u^l_1}+u^l_3\frac{\pa}{\pa u^l_2}+\dots\right)\right)-\\
&-\left(b^i_1+D_x\right)\left(\frac{\pa}{\pa u^j_1}+b^j_1\frac{\pa}{\pa u^j_2}+b^j_2\frac{\pa}{\pa u^j_3}+\dots\right)=\\
&=X_j(b^i_1)+X_j(u^j_2)\frac{\pa}{\pa u^j_1}+X_j(u^j_3)\frac{\pa}{\pa u^j_2}+\dots-D_x(b^j_1)\frac{\pa}{\pa u^j_2}-D_x(b^j_2)\frac{\pa}{\pa u^j_3}-\dots=\\
&=a_{ij}+b^j_1\frac{\pa}{\pa u^j_1}+b^j_2\frac{\pa}{\pa u^j_2}+\dots+(b^j_1 b^j_1-b^j_2)\frac{\pa}{\pa u^j_2}+(b^j_1 b^j_2-b^j_3)\frac{\pa}{\pa u^j_3}+\dots=\\
&=a_{ij}+b^j_1\left(\frac{\pa}{\pa u^j_1}+b^j_1\frac{\pa}{\pa u^j_2}+b^j_2\frac{\pa}{\pa u^j_3}+\dots\right)=a_{ij}+b^j_1 X_j.\quad\Box
\end{align*}
\begin{proposition}\label{proprelk}
Operators $X_j$ satisfy the relation
$$
X_j^k (D_x+b^i_1)=(D_x+b^i_1+k b^j_1)X_j+\left(k a_{ij}+\frac{k(k-1)}{2}a_{jj}\right)X_j^{k-1}
$$
for all $i,j=1,\dots,r$ and for all natural $k$.
\end{proposition}

{\bf Proof}. Use induction in $k$. The base of induction follows from Proposition~\ref{proprel}. $\Box$
\begin{proposition}
Let $k_1,\dots,k_s$ be arbitrary natural numbers and assume $1\leqslant j_1<j_2<\dots<j_s\leqslant r$. Then
\begin{multline}
\label{aee}
X^{k_s}_{j_s} X^{k_{s-1}}_{j_{s-1}}\dots X^{k_1}_{j_1}\left(D_x+b^i_1\right)=\left(D_x+k_s b^{j_s}_1+\dots+k_1 b^{j_1}_1+b^i_1\right)
X^{k_s}_{j_s} X^{k_{s-1}}_{j_{s-1}}\dots X^{k_1}_{j_1}+\\
+k_s\left(a_{ij_s}+\frac{k_s-1}{2}a_{j_sj_s}+k_{s-1}a_{j_{s-1}j_s}+k_{s-2}a_{j_{s-2}j_s}+\dots+k_1 a_{j_1 j_s}\right)X^{k_s-1}_{j_s} X^{k_{s-1}}_{j_{s-1}}\dots X^{k_1}_{j_1}+\\
+k_{s-1}\left(a_{ij_{s-1}}+\frac{k_{s-1}-1}{2}a_{j_{s-1}j_{s-1}}+k_{s-2}a_{j_{s-2}j_{s-1}}+k_{s-3}a_{j_{s-3}j_{s-1}}+\dots+k_1 a_{j_1 j_{s-1}}\right)X^{k_s}_{j_s} X^{k_{s-1}-1}_{j_{s-1}}\dots X^{k_1}_{j_1}+\dots\\
+k_3\left(a_{ij_3}+\frac{k_3-1}{2}a_{j_3j_3}+k_2 a_{j_2 j_3}+k_1 a_{j_1j_3}\right)X^{k_s}_{j_s}\dots X^{k_3-1}_{j_3}X^{k_2}_{j_2}X^{k_1}_{j_1}+\\
+k_2\left(a_{ij_2}+\frac{k_2-1}{2}a_{j_2j_2}+k_1 a_{j_1j_2}\right)X^{k_s}_{j_s}\dots X^{k_2-1}_{j_2}X^{k_1}_{j_1}+
k_1\left(a_{ij_1}+\frac{k_1-1}{2}a_{j_1j_1}\right)X^{k_s}_{j_s}\dots X^{k_2}_{j_2}X^{k_1-1}_{j_1}
\end{multline}
for any $i=1,\dots,r$.
\end{proposition}

{\bf Proof}. Use induction in $s$. The base of induction follows from Proposition~\ref{proprelk}. $\Box$

\begin{proposition}
Let $k\in\mathbb N$. Then $D_x^k (t_i)$ is a polynomial in $t_i,\dots,t_r$ of the degree $k+1$,
\begin{equation}
\label{acc}
D^k_x(t_i)=\sum\limits_{\varkappa=1}^{k+1} C^k_{k_i\dots k_r} t^{k_i}_i t^{k_{i+1}}_{i+1}\dots t^{k_r}_r,
\end{equation}
where the summation is taken over all partitions $\varkappa=k_i+\dots+k_r$ into a sum of non-negative whole numbers such that $k_i>0$. Coefficients are given by
\begin{equation}
\label{ajj}
C^k_{k_i\dots k_r}=\frac{1}{k_i !\dots k_r !}X^{k_r}_r X^{k_{r-1}}_{r-1}\dots X^{k_{i+1}}_{i+1}X^{k_i-1}_i(b^k_i).
\end{equation}
\end{proposition}

{\bf Proof}. Use induction in $k$. The base of the induction follows from~(\ref{add}) since $X_j (b_1^i)=a_{ij}$. Assume the proposition holds for
$$
D_x (t_i),\quad D^2_x (t_i),\dots, D^k_x(t_i)
$$
and differentiate~(\ref{acc}) with respect to $x$. The total derivative of the coefficient $C^k_{k_i\dots k_s}$ contributes only to the coefficient of
$t^{k_i}_i\dots t^{k_r}_r$ in the expansion of $D^{k+1}_x$. Due to~(\ref{add}), the derivative of $t^{k_i}_i\dots t^{k_r}_r$ contributes to the coefficient of
$t^{k_i}_i\dots t^{k_r}_r$ and to all coefficients of the terms of degree $k_i+\dots+k_r+1$ such that only one power differs from $k_i,\dots,k_r$ by one.
First consider one of the leading coefficients $C^{k+1}_{k_i\dots k_r}$ in $D^{k+1}_x (t_i)$. It appears as the result of differentiation of the terms
$$
t^{k_i-1}_i t^{k_{i+1}}_{i+1}\dots t^{k_r}_r,\quad t^{k_i}_i t^{k_{i+1}-1}_{i+1}\dots t^{k_r}_r,\dots,t^{k_i}_i t^{k_{i+1}}_{i+1}\dots t^{k_r-1}_r,
$$
where the corresponding power $k_s-1$ is non-negative. Hence,
\begin{align}
\notag
C^{k+1}_{k_i k_{i+1}\dots k_r}&=C^k_{k_i-1,k_{i+1}\dots k_r}(k_i-1)\frac{a_{ii}}{2}+
C^k_{k_i k_{i+1}-1\dots k_r}\left(k_i a_{i,i+1}+(k_{i+1}-1)\frac{a_{i+1,i+1}}{2}\right)+\\
\notag
&+C^k_{k_i k_{i+1} k_{i+2}-1\dots k_r}\left(k_i a_{i,i+2}+k_{i+1} a_{i+1,i+2}+(k_{i+2}-1)\frac{a_{i+2,i+2}}{2}\right)+\dots+\\
\notag
&+C^k_{k_i k_{i+1}-1\dots k_r-1}\left(k_i a_{i,r}+k_{i+1} a_{i+1,r}+k_{i+2} a_{i+2,r}+\dots+k_{r-1} a_{r-1,r}+(k_r-1)\frac{a_{rr}}{2}\right)=\\
\notag
&=\frac{1}{k_i !\dots k_r !}\left(k_i\frac{k_i-1}{2}a_{ii}X^{k_r}_r\dots X^{k_{i+1}}_{i+1}X^{k_i-2}_i(b^k_i)+\right.\\
\notag
&+k_{i+1}\left(k_i a_{i,i+1}+\frac{k_{i+1}-1}{2}a_{i+1,i+1}\right)X^{k_r}_r\dots X^{k_{i+1}-1}_{i+1}X^{k_i-1}_i(b^k_i)+\dots+\\
\label{aff}
&+\left.k_r\left(k_i a_{i,r}+k_{i+1} a_{i+1,r}+k_{i+2} a_{i+2,r}+\dots+k_{r-1} a_{r-1,r}+\frac{k_r-1}{2}a_{rr}\right)
X^{k_r-1}_r\dots X^{k_{i+1}}_{i+1}X^{k_i-1}_i(b^k_i)\right).
\end{align}
Apply now formula~(\ref{aee}):
\begin{multline}
\label{agg}
X^{k_r}_r X^{k_{r-1}}_{r-1}\dots X^{k_{i+1}}_{i+1}X^{k_i-1}_i(b^{k+1}_i)=X^{k_r}_r X^{k_{r-1}}_{r-1}\dots X^{k_{i+1}}_{i+1}X^{k_i-1}_i(D_x+b^i_1)(b^k_i)=\\
=k_r\left(a_{ir}+\frac{k_r-1}{2}a_{rr}+k_{r-1}a_{r-1,r}+\dots+(k_i-1) a_{ir}\right)X^{k_r-1}_r X^{k_{r-1}}_{r-1}\dots X^{k_i-1}_i(b^k_i)+\dots+\\
+k_{i+1}\left(a_{i,i+1}+\frac{k_{i+1}-1}{2}a_{i+1,i+1}+(k_i-1) a_{i,i+1}\right)X^{k_r}_r\dots X^{k_{i+1}-1}_{i+1} X^{k_i-1}_i(b^k_i)+\\
+k_i\left(a_{ii}+\frac{k_i-2}{2}a_{ii}\right)X^{k_r}_r\dots X^{k_{i+1}}_{i+1}X^{k_i-2}_i(b^k_i).
\end{multline}
Here we used relation
$$
X^{k_r}_r X^{k_{r-1}}_{r-1}\dots X^{k_{i+1}}_{i+1}X^{k_i-1}_i(b^{k}_i)=0,
$$
which holds since $k_i+\dots+k_r=k+2$. Comparing formulas~(\ref{aff}) and~(\ref{agg}) proves the claim for leading coefficients in the expansion for
$D^{k+1}_x (t_i)$.

The formula for non-leading coefficients is proved similarly, although the calculation is more nasty in this case since one has to take into account the terms
$b^s_1 t^{k_i}_i t^{k_{i+1}}_{i+1}\dots t^{k_r}_r$, where $s=i,\dots r$, that come from the differentiation of $t^{k_i}_i t^{k_{i+1}}_{i+1}\dots t^{k_r}_r$,
and the terms that appear from the differentiation of the coefficient $C^k_{k_i\dots k_s}$ which are non-zero if $\varkappa<k+1$. Note that
formula~(\ref{aee}) has to be used also to simplify the coefficient that comes from differentiation of $C^k_{k_i\dots k_s}$. $\Box$

\begin{proposition}\label{proplead}
Polynomial $D_x^k(t_i)$ is divisible by $t_i$ for all $i=1,\dots,r$ and $k\in\mathbb N$. The coefficient of $t_i$ in $D_x^k(t_i)$ equals $b^k_i$:
$$
D_x^k(t_i)=t_i(b^k_i+\dots).
$$
\end{proposition}

{\bf Proof}. The first claim follows from~(\ref{add}). The second claim immediately follows from~(\ref{ajj}). $\Box$

\begin{proposition}
Let $I$ be an analytic $y$-integral of exponential system~(\ref{expsyst}). Then
\begin{equation}
\label{ahh}
I_{n+1}-I_n=\sum\limits_{m=1}^\infty \left(\sum\limits_{k_1+\dots+k_r=m}^{} \frac{1}{k_1!\dots k_r!}
X^{k_r}_r X^{k_{r-1}}_{r-1}\dots X^{k_1}_1 (I) t_1^{k_1}\dots t^{k_r}_r\right),
\end{equation}
where the sum is taken over all partitions $m=k_1+\dots+k_r$ such that $k_1,\dots, k_r\geqslant 0$.
\end{proposition}

{\bf Proof}. Use induction in $m$. If $m=1$, then we need to prove that the coefficient of $t_i$ in the sum~(\ref{abb}) equals $X_i (I)$ for all $i=1,\dots,r$.
Since the monomial $t_i$ is contained only in the expansions of form~(\ref{acc}) for $t_i,t_{i,x},t_{i,xx},\dots$, the coefficient of $t_i$ in~(\ref{abb}) equals
$$
1\cdot\frac{\pa I}{\pa u^i_1}+C^1_{1,0,\dots,0}\frac{\pa I}{\pa u^i_2}+C^2_{1,0,\dots,0}\frac{\pa I}{\pa u^i_3}+\dots=
\frac{\pa I}{\pa u^i_1}+b^i_1\frac{\pa I}{\pa u^i_2}+b^i_2\frac{\pa I}{\pa u^i_3}+\dots=X_i (I)
$$
due to Proposition~\ref{proplead}.

Assume now the coefficient $B$ of $t_l^{k_l}\dots t_j^{k_j}$ in~(\ref{abb}) has the form
$$
\frac{1}{k_l!\dots k_j!} X^{k_j}_j X^{k_{j-1}}_{j-1}\dots X^{k_l}_l (I)
$$
for all $l<j$ and $k_l+\dots+k_j\leqslant m$. According to~(\ref{abb}), the coefficient $B$ is obtained from the product
$$
\prod\limits_{i=1}^r\prod\limits_{k=1}^d\left(D^{k-1}_x(t_i)\right)^{\lambda_{ik}}
$$
by extracting the coefficients of $t_l^{k_l}\dots t_j^{k_j}$ and multiplying them by appropriate multiple derivatives of $I$. Hence,
$$
B=\sum\limits_p^{}\left(\left(\mu_p\prod\limits_{i,k}^{} C^k_{i,q_l\dots q_j}\right)\frac{\pa^s I}{(\pa u^1_1)^{\lambda_{11}}\dots(\pa u^r_d)^{\lambda_{rd}}}\right),
$$
where $C^k_{i,q_l\dots q_j}$ are the coefficients of the form~(\ref{ajj}) in one of the expansions of $D^k_x (t_i)$ in~(\ref{abb}), $0\leqslant q_i\leqslant k_i$
for all $i=l,\dots,j$ (here $q_l\dots q_j$ depend on $i$ and $k$) and $\mu_p\in\mathbb R$ are products of binomial coefficients which appear when expansions of
$D^k_x (t_i)$ are raised to powers $\lambda_{i,k}$ in~(\ref{abb}). We do not need to specify the set of indices over which the sum and the products are taken ---
we will only need the general form of $B$.

Consider the coefficient $\tilde B$ of $t_l^{k_l}\dots t_j^{k_j}t_{j+1}$ in~(\ref{abb}). It is the sum of the terms of two types: the first one
is the product of the coefficient of $t_l^{k_l}\dots t_j^{k_j}$ in~(\ref{abb}) by the coefficient of $t_{j+1}$ in expansions of
$D^k_x (t_{j+1})$ for all $k=0,1,\dots$, and the second one appears when $t_{j+1}$ is additionally taken from one of the expansions of $D^k(t_i)$, where $i<j+1$.
For the terms of the first type, the derivative of the function $I$ in~(\ref{abb}) is differentiated by $u^{j+1}_{k+1}$ since the whole expression is multiplied by
$D^k_x (t_{j+1})$. According to Proposition~\ref{proplead} the coefficient of $t_{j+1}$ in $D^k_x (t_{j+1})$ equals $b^{j+1}_k$. Therefore, it follows from the
inductive assumption that the contribution of the terms of the first type to $\tilde B$ has the form
\begin{multline}
\label{akk}
\frac{1}{k_l!\dots k_j!}\left(X^{k_j}_j X^{k_{j-1}}_{j-1}\dots X^{k_l}_l \left(\frac{\pa I}{\pa u^{j+1}_1}\right)+
b^{j+1}_1 X^{k_j}_j X^{k_{j-1}}_{j-1}\dots X^{k_l}_l \left(\frac{\pa I}{\pa u^{j+1}_2}\right)+\right.\\
+\left.b^{j+1}_2 X^{k_j}_j X^{k_{j-1}}_{j-1}\dots X^{k_l}_l \left(\frac{\pa I}{\pa u^{j+1}_3}\right)+\dots\right).
\end{multline}

Examine now the contribution of the terms of the second type to $\tilde B$. Since $t_{j+1}$ is additionally taken from one of the expansions of $D^k(t_i)$ in this
case, the terms contributing to $\tilde B$ have the form
$$
\left(\mu_p\prod\limits_{i,k}^{} C^k_{i,q_l\dots q_j q_{j+1}}\right)\frac{\pa^s I}{(\pa u^1_1)^{\lambda_{11}}\dots(\pa u^r_d)^{\lambda_{rd}}},
$$
where $q_{j+1}=1$ for only one pair $(i,k)$ and is zero for all other pairs $(i,k)$. Note that due to~(\ref{ajj})
\begin{equation}
\label{amm}
C^k_{i,q_l\dots q_j q_{j+1}}=X_{j+1}\left(C^k_{i,q_l\dots q_j}\right)
\end{equation}
for all $i=1,\dots,r$ and for all $k\in\mathbb N$.

Since $X^{k_j}_j X^{k_{j-1}}_{j-1}\dots X^{k_l}_l (I)$ is a linear combination of various multiple derivatives of the function $I$ with certain coefficients and
$X_{j+1}$ is a linear first order operator, all the terms in the expression for
$$
X_{j+1}\left(\frac{1}{k_l!\dots k_j!} X^{k_j}_j X^{k_{j-1}}_{j-1}\dots X^{k_l}_l (I)\right)
$$
can be grouped into two families: the operator $X_{j+1}$ is applied to the derivatives of $I$ in the first family and it is applied to the coefficients in the
second family. Clearly, it follows from~(\ref{akk},\ref{amm}) and the Leibniz rule that the first family corresponds to the terms of the first type in $\tilde B$
and the second family corresponds to the terms of the second type. Therefore, $\tilde B=X_{j+1}(B)$.

The proof that the coefficient of $t_l^{k_l}\dots t_j^{k_j+1}$ in~(\ref{abb}) has the form
$$
\frac{1}{k_l!\dots k_{j-1}! (k_j+1)!} X^{k_j+1}_j X^{k_{j-1}}_{j-1}\dots X^{k_l}_l (I)
$$
is conducted in the same way, but one has to take into account additional coefficient $k_j+1$ that comes from multiplicity in this case. $\Box$

Proof of Theorem~\ref{thmmain} immediately follows from~(\ref{ahh}) since every $y$-integral annihilates the whole characteristic algebra and, in particular,
it annihilates the generators $\tilde X_1,\dots\tilde X_r$. Since $X_i=e^{-w^i}\tilde X_i$, it also annihilates operators $X_1,\dots X_r$. $\Box$

\section{Characteristic algebra of exponential system}\label{charcont}

In this Section we review a number of basic properties of characteristic algebras for exponential systems corresponding to the Cartan matrices of simple Lie algebras in
the continuous case.

\begin{lemma}\label{shlemma}
Let
$$
X=\sum\limits_{i=1}^r\sum\limits_{k=1}^\infty P^i_k\frac{\pa}{\pa u^i_k},\quad D=\sum\limits_{i=1}^r\sum\limits_{k=0}^\infty u^i_{k+1}\frac{\pa}{\pa u^i_k},
$$
where $P^i_k=P^i_k(u^1,\dots,u^r,u^1_1,\dots,u^r_1,u^1_2,\dots,u^r_2,\dots)$ and $u^i_0=u^i$. If $[D,X]=0$, then $X=0$.
\end{lemma}

This lemma and its various analogs are widely used for explicit description of characteristic algebras. The proof is straightforward and trivial.

\begin{proposition}{\rm\cite{ShJa}}
For any $1\leqslant i_1,\dots,i_k\leqslant r$ operators $X_i$ satisfy following commutation relations:
\begin{multline}
\label{comm}
[D,[X_{i_1},[X_{i_2},\dots,[X_{i_{k-1}},X_{i_k}]\dots]]]=-(b^{i_1}_1+\dots+b^{i_k}_1)[X_{i_1},[X_{i_2},\dots,[X_{i_{k-1}},X_{i_k}]\dots]]-\\
-(a_{i_2 i_1}+a_{i_3 i_1}+\dots+a_{i_k i_1})[X_{i_2},[X_{i_3},\dots,[X_{i_{k-1}},X_{i_k}]\dots]]-\\
-(a_{i_3 i_2}+a_{i_4 i_2}+\dots+a_{i_k i_2})[X_{i_1},[X_{i_3},\dots,[X_{i_{k-1}},X_{i_k}]\dots]]-\dots\\
-(a_{i_{k-1} i_{k-2}}+a_{i_k i_{k-2}})[X_{i_1},\dots,[X_{i_{k-3}},[X_{i_{k-1}},X_{i_k}]]\dots]-\\
-a_{i_k i_{k-1}}[X_{i_1},\dots,[X_{i_{k-3}},[X_{i_{k-2}},X_{i_k}]]\dots]
+a_{i_{k-1} i_k}[X_{i_1},\dots,[X_{i_{k-3}},[X_{i_{k-2}},X_{i_{k-1}}]]\dots].
\end{multline}
\end{proposition}

Technical Propositions~\ref{propa}--\ref{propx} can be proved straightforwardly using Lemma~\ref{shlemma}.
\begin{proposition}\label{propa}
Reduced characteristic algebra of exponential system~(\ref{expsyst}) corresponding to the $A$-series Cartan matrix of the rank $r$ is a linear span
of the following vector fields:
\begin{align}
&X_k,\quad k=1,2,\dots,r,
\notag\\
&[X_k,[X_{k+1},\dots,[X_{l-1},X_l]\dots]],\quad 1\leqslant k<l\leqslant r.
\notag
\end{align}
\end{proposition}
\begin{proposition}\label{propb}
Reduced characteristic algebra of exponential system~(\ref{expsyst}) corresponding to the $B$-series Cartan matrix of the rank $r$ is a linear span
of the following linearly independent vector fields:
\begin{align}
&X_k,\quad k=1,2,\dots,r,
\notag\\
&[X_k,[X_{k+1},\dots,[X_{l-1},X_l]\dots]],\quad 1\leqslant k<l\leqslant r,
\notag\\
&[X_k,[X_{k-1},[X_{k-2},\dots,[X_2,[X_1,[X_1,[X_2,\dots,[X_{l-2},[X_{l-1},X_l]]\dots]]]]\dots]]],\quad 1\leqslant k<l\leqslant r.
\notag
\end{align}
\end{proposition}
\begin{proposition}\label{propc}
Reduced characteristic algebra of exponential system~(\ref{expsyst}) corresponding to the $C$-series Cartan matrix of the rank $r$ is a linear span
of the following linearly independent vector fields:
\begin{align}
&X_k,\quad k=1,2,\dots,r,
\notag\\
&[X_k,[X_{k+1},\dots,[X_{l-1},X_l]\dots]],\quad 1\leqslant k<l\leqslant r,
\notag\\
&[X_k,[X_{k-1},[X_{k-2},\dots,[X_2,[X_1,[X_2,\dots,[X_{l-2},[X_{l-1},X_l]]\dots]]]\dots]]],\quad 2\leqslant k\leqslant l\leqslant r.
\notag
\end{align}
\end{proposition}
\begin{proposition}\label{propd}
Reduced characteristic algebra of exponential system~(\ref{expsyst}) corresponding to the $D$-series Cartan matrix of the rank $r$ is a linear span
of the following linearly independent vector fields:
\begin{align}
&X_k,\quad k=1,2,\dots,r,
\notag\\
&[X_k,[X_{k+1},\dots,[X_{l-1},X_l]\dots]],\quad 2\leqslant k<l\leqslant r,
\notag\\
&[X_1,[X_2,\dots,[X_{l-1},X_l]\dots]],\quad 3\leqslant l\leqslant r,
\notag\\
&[X_1,[X_3,\dots,[X_{l-1},X_l]\dots]],\quad 3\leqslant l\leqslant r,
\notag\\
&[X_k,[X_{k-1},[X_{k-2},\dots,[X_3,[X_1,[X_2,[X_3,\dots,[X_{l-2},[X_{l-1},X_l]]\dots]]]]\dots]]],\quad 3\leqslant k<l\leqslant r.
\notag
\end{align}
\end{proposition}
\begin{remark}
\rm
The above bases for the series $A$ and $D$ can be found in~\cite{ShJa} but with no proofs. Detailed proofs for all these four cases can be found in~\cite{Pe}.
\end{remark}
\begin{remark}
\rm
One can verify that in characteristic algebras of exponential systems corresponding to the Cartan matrices of the series $A$--$D$ there are no non-trivial
relations between multiple commutators of the form $[X_{i_1},[X_{i_2},\dots,X_{i_k}]]$. More precisely, any such non-zero commutator is proportional to some
element from the corresponding basis (see Propositions~\ref{propa}--\ref{propd}).
\end{remark}
\begin{proposition}\label{propx}
Reduced characteristic algebra of exponential systems~(\ref{expsyst}) corresponding to the Cartan matrices of the $E$-series, of $G_2$ and $F_4$ root systems are
finite-dimensional, and in each of these cases there exist a basis such that any non-zero multiple commutator $[X_{i_1},[X_{i_2},\dots,X_{i_k}]]$ is proportional
to some element from this basis\footnote{We do not provide here suitable bases because they cannot be represented in a compact form like in
Propositions~\ref{propa}--\ref{propd}.}.
\end{proposition}

\section{Characteristic $x$-integrals of semi-discrete exponential systems}\label{xint}

Integral preserving discretization~(\ref{sdexpsyst}) of a Darboux integrable exponential system~(\ref{expsyst}) admits a complete family of independent $n$-integrals.
Darboux integrability of a semi-discrete system also requires the existence of a complete family of essentially independent $x$-integrals, and since the variables $x$ and $n$ do not
enter the equations symmetrically (one of them is continuous and the other one is discrete), such family of integrals cannot be obtained just by renaming the
variables (as in the continuous case). Complete family of characteristic integrals was obtained explicitly in~\cite{Sm15} for semi-discrete $A$- and $C$-series lattices, but
this approach is not applicable for the lattices associated with the Cartan matrices of other simple Lie algebras since it is not known whether they are reductions
of an $A$-series lattice or not. In this section we are going to prove the existence of complete families of characteristic $x$-integrals for semi-discrete lattices
corresponding to the Cartan matrices of all simple Lie algebras using the concept of characteristic algebra.

Characteristic algebras for (semi)-discrete hyperbolic system were defined in~\cite{Ha1}--\cite{HSZ}; these Lie algebras for (semi)-discrete exponential
systems~(\ref{sdexpsyst}),(\ref{pdexpsyst}) associated with the Cartan matrices of rank $2$ where described explicitly in terms of generators and relations
in~\cite{HZY,GHY}. Characteristic algebras are an effective tool in the study of Darboux integrability of hyperbolic equations both in the continuous and in
the (semi)-discrete cases (see papers~\cite{HPo}--\cite{HK21} by Ufa mathematical school).

In this section we define a special Lie algebra of differential operators that controls the existence of complete family of $x$-integrals
for semi-discrete exponential systems~(\ref{sdexpsyst}). Although this Lie algebra is very similar by its properties to characteristic algebra defined in~\cite{HP}
and its construction is based on the same ideas, these Lie algebras are not isomorphic.

We will not describe the construction of characteristic algebra in the semi-discrete case from~\cite{HP} here. Instead, we introduce a Lie algebra with
similar properties that allows to prove Darboux integrability of all exponential systems associated with the Cartan matrices of simple Lie algebras.

Consider semi-discrete exponential system~(\ref{sdexpsyst}) and introduce the following notation:
$$
v^i_n=u^i_{n+1}-u^i_n,\quad w^i_n=a_{i1}u^1_n+\dots+a_{ir}u^r_n,\quad z^i_n=w^i_{n+1}-w^i_n,\quad
\Delta^i_n=\exp\left(\frac{a_{ii}}{2}v^i_n+a_{i,i+1}v^{i+1}_n+\dots+a_{ir} v^r_n\right),
$$
where $i=1,\dots,r$. Hence, equations~(\ref{sdexpsyst}) can be rewritten as
\begin{equation}
\label{sdmod}
v^i_{n,x}=e^{w^i_n}\Delta^i_n,\quad i=1,\dots,r.
\end{equation}
Consider differential operators
$$
Y_i=c^i_0\frac{\pa}{\pa v^i_n}+c^i_1\frac{\pa}{\pa v^i_{n+1}}+c^i_2\frac{\pa}{\pa v^i_{n+2}}+\dots,\quad i=1,\dots,r,
$$
where
\begin{equation}
\label{ann}
c^i_0=\Delta^i_n,\quad c^i_1=\exp\left(z^i_n\right)\Delta^i_{n+1},\dots,c^i_k=\exp\left(z^i_n+z^i_{n+1}+\dots+z^i_{n+k-1}\right)\Delta^i_{n+k},\quad i=1,\dots,r.
\end{equation}
\begin{proposition}
Let the matrix $M=(a_{ij})$ of exponential system~(\ref{sdexpsyst}) be non-degenerate. Then function
$$
J=J(v^1_n,\dots,v^r_n,v^1_{n+1},\dots,v^r_{n+1},v^1_{n+2},\dots,v^r_{n+2},\dots)
$$
is an $x$-integral of~(\ref{sdexpsyst}) if and only if it annihilates operators $Y_1,\dots,Y_r$.
\end{proposition}
{\bf Proof}. Since
$$
v^i_{n+1,x}=e^{w^i_{n+1}}\Delta^i_{n+1}=e^{w^i_n}\exp(z^i_n)\Delta^i_{n+1}
$$
due to~(\ref{sdmod}), one can easily verify by induction that
$$
v^i_{n+k,x}=e^{w^i_n}\exp(z^i_n+z^i_{n+1}+\dots+z^i_{n+k-1})\Delta^i_{n+k}.
$$
Hence, apply the total derivative with respect to $x$:
\begin{multline*}
D_x(J)=\sum\limits_{i=1}^r\sum\limits_{k=0}^\infty v^i_{n+k,x}\frac{\pa J}{\pa v^i_{n+k}}=
\sum\limits_{i=1}^r\sum\limits_{k=0}^\infty e^{w^i_n}\exp(z^i_n+z^i_{n+1}+\dots+z^i_{n+k-1})\Delta^i_{n+k}\frac{\pa J}{\pa v^i_{n+k}}=\\
=\sum\limits_{i=1}^r\sum\limits_{k=0}^\infty e^{w^i_n} c^i_k\frac{\pa J}{\pa v^i_{n+k}}=
\sum\limits_{i=1}^r e^{w^i_n} Y_i (J).
\end{multline*}
Therefore, if $M$ is non-degenerate, then exponents $e^{w^i_n}$ are linearly independent and $J$ is an $x$-integral if and only if it annihilates
operators $Y_1,\dots,Y_r$. $\Box$

Variables
$$
u^1_n,\dots,u^r_n,v^1_n,\dots,v^r_n,v^1_{n+1},\dots,v^r_{n+1},v^1_{n+2},\dots,v^r_{n+2},\dots
$$
are independent. Lie algebra $\tilde{\mathcal L}$ generated by vector fields
$$
\frac{\pa}{\pa u^1_n},\dots,\frac{\pa}{\pa u^r_n},\ \tilde Y_1,\dots,\tilde Y_r,
$$
where $\tilde Y_i=e^{w^i_n}Y_i$ for all $i=1,\dots,r$, is a discrete analog of the characteristic algebra for exponential
system~(\ref{expsyst}). Similarly, Lie algebra $\mathcal L$ generated by vector fields $Y_1,\dots,Y_r$ is a discrete analog of the reduced
characteristic algebra for~(\ref{expsyst}). Nevertheless, in order to avoid ambiguity in the use of terminology, we will call this Lie algebra
{\it the defining algebra} for semi-discrete system~(\ref{sdexpsyst}). The following theorem is proved exactly as Theorem~\ref{thchar} in the continuous case.
\begin{theorem}\label{thsdchar}
Semi-discrete exponential system~(\ref{sdexpsyst}) admits a complete family of essentially independent $x$-integrals if and only if its defining algebra is
finite-dimensional.
\end{theorem}

We are going to prove the existence of a complete family of independent $x$-integrals for semi-discrete exponential systems~(\ref{sdexpsyst}) corresponding to
the Cartan matrices of all simple Lie algebras by applying Theorem~\ref{thsdchar} and by describing the defining algebras $\mathcal L$ explicitly.

\begin{proposition}
Let $T$ be the shift operator, $T(v_n)=v_{n+1}$. Then
\begin{multline}
T [Y_{i_1},[Y_{i_2},\dots,[Y_{i_{k-1}},Y_{i_k}]\dots]]T^{-1}=\exp(-(z^{i_1}_n+\dots+z^{i_k}_n))\left([Y_{i_1},[Y_{i_2},\dots,[Y_{i_{k-1}},Y_{i_k}]\dots]]-\right.\\
-c^{i_1}_n(a_{i_2 i_1}+a_{i_3 i_1}+\dots+a_{i_k i_1})[Y_{i_2},[Y_{i_3},\dots,[Y_{i_{k-1}},Y_{i_k}]\dots]]-\\
-c^{i_2}_n(a_{i_3 i_2}+a_{i_4 i_2}+\dots+a_{i_k i_2})[Y_{i_1},[Y_{i_3},\dots,[Y_{i_{k-1}},Y_{i_k}]\dots]]-\dots\\
-c^{i_{k-2}}_n(a_{i_{k-1} i_{k-2}}+a_{i_k i_{k-2}})[Y_{i_1},\dots,[Y_{i_{k-3}},[Y_{i_{k-1}},Y_{i_k}]]\dots]-\\
-c^{i_{k-1}}_n a_{i_k i_{k-1}}[Y_{i_1},\dots,[Y_{i_{k-3}},[Y_{i_{k-2}},Y_{i_k}]]\dots]
\left. +c^{i_k}_n a_{i_{k-1} i_k}[Y_{i_1},\dots,[Y_{i_{k-3}},[Y_{i_{k-2}},Y_{i_{k-1}}]]\dots]\right)+\dots,
\label{app}
\end{multline}
where the dots in the end stand for multiple commutators of the degrees $1,\dots,k-2$.
\end{proposition}
{\bf Proof}

It follows from relations~(\ref{ann}) that
$$
T Y_i T^{-1}=e^{-z^i_n} Y_i
$$
for all $i=1,\dots, r$. Formula~(\ref{app}) is proved by induction using relation
$$
Y_{i_0}\left(\exp(-(z^{i_1}_n+\dots+z^{i_k}_n))\right)=-c^{i_0}_n(a_{i_1 i_0}+a_{i_2 i_0}+\dots+a_{i_k i_0})
$$
and representation
\begin{multline*}
T[Y_{i_0},[Y_{i_1},[Y_{i_2},\dots,[Y_{i_{k-1}},Y_{i_k}]\dots]]]T^{-1}=\\
=\left(TY_{i_0} T^{-1}\right)\left(T [Y_{i_1},[Y_{i_2},\dots,[Y_{i_{k-1}},Y_{i_k}]\dots]]T^{-1}\right)-
\left(T [Y_{i_1},[Y_{i_2},\dots,[Y_{i_{k-1}},Y_{i_k}]\dots]]T^{-1}\right)\left(TY_{i_0} T^{-1}\right).\quad\Box
\end{multline*}
\begin{theorem}
The defining algebra of semi-discrete exponential system~(\ref{sdexpsyst}) corresponding to the Cartan matrix of any simple Lie algebra is finite-dimensional.
\end{theorem}

{\bf Proof}. We will prove that the defining algebra of semi-discrete exponential system~(\ref{sdexpsyst}) corresponding to the Cartan matrix of any simple
Lie algebra is isomorphic to the reduced characteristic algebra of its continuous analog. Since in characteristic algebras corresponding to the Cartan matrices
of all simple Lie algebras there are no relations between non-trivial multiple commutators except for the skew-symmetry, the Jacobi identity and their implications
(see Propositions~\ref{propa}--\ref{propx}), it follows from Lemma~\ref{shlemma} and from formula~(\ref{comm}) that
commutator
\begin{equation}
\label{aqq}
[X_{i_1},[X_{i_2},\dots,[X_{i_{k-1}},X_{i_k}]\dots]]
\end{equation}
is trivial if and only if all terms in~(\ref{comm}) vanish. Therefore, either commutator
$$
[X_{i_1},[X_{i_2},\dots,[X_{l-1},[X_{l+1},\dots,[X_{i_{k-1}},X_{i_k}]\dots]]\dots]]
$$
is trivial, or its coefficient
$$
a_{i_{l+1} i_l}+a_{i_{l+2} i_l}+\dots+a_{i_k i_l}
$$
is zero for all $l=1,\dots,k$.

For each of the Cartan matrices the isomorphism is established by induction in the degree of multiple commutators. Note that the coefficients of multiple
commutators of degree $k-1$ in formula~(\ref{app}) are the same as the coefficients in~(\ref{comm}). If
commutator~(\ref{aqq}) is trivial, then all the terms in~(\ref{comm}) vanish. Hence, relation~(\ref{app}) for the corresponding multiple commutator
of $Y_{i_1},\dots, Y_{i_k}$ does not contain terms of degree $k-1$. Careful analysis of the relations between coefficients in formula~(\ref{app}) for
multiple commutators of degrees $l$ and $l-1$ for each Cartan matrix shows that the absence of the terms of degree $k-1$ yields the absence of all terms of
degrees $1,\dots,k-2$. Hence, commutator~(\ref{aqq}) satisfies relation
$$
T [Y_{i_1},[Y_{i_2},\dots,[Y_{i_{k-1}},Y_{i_k}]\dots]]T^{-1}=\exp(-(z^{i_1}_n+\dots+z^{i_k}_n))[Y_{i_1},[Y_{i_2},\dots,[Y_{i_{k-1}},Y_{i_k}]\dots]]
$$
which holds only if $k=1$. Therefore, if commutator~(\ref{aqq}) is trivial in the reduced characteristic algebra of exponential system in the continuous case, then the
corresponding multiple commutator of $Y_{i_1},\dots, Y_{i_k}$ also vanishes. This proves the isomorphism. $\Box$

\begin{remark}
\rm
We used the concept of characteristic algebra in order to prove the existence of a complete family of $x$-integrals. Although our attempts to find explicit
formulas for $x$-integrals did not give any result in general, some integrals can be found explicitly. For example, semi-discrete Toda lattice~(\ref{sdexpsyst})
corresponding to the $B$-series Cartan matrix of the rank $r\geqslant 3$ admits $x$-integral
\begin{multline*}
J=\sum_{j=2}^{r-1}\left(\exp(v^{j+1}_{n+j+1}-v^j_{n+j+1})+\exp(v^i_n-v^{j+1}_{n+1})\right)+\exp(v^2_{n+1}-2v^1_{n+1})+\\
+\exp(2v^1_n-v^2_{n+1})+2\exp(v^1_n-v^1_{n+1})+\exp(v^r_n)+\exp(-v^r_{n+r}).
\end{multline*}
\end{remark}
\begin{remark}
\rm
It would have been interesting to prove that Habibullin's integral preserving discretization leads to Darboux integrable systems in the purely discrete case
as well. Entirely discrete exponential systems were introduced in~\cite{GHY} and it was proved there that for all Cartan matrices of the rank $2$ characteristic
$x$-integrals of corresponding semi-discrete lattices appear to be $m$-integrals of their purely discrete analogs. It was conjectured in~\cite{GHY} that the same
property holds for purely discrete exponential systems associated with the Cartan matrices of all simple Lie algebras, but unfortunately our approach appears to be
not applicable for proving the integral preserving property in the entirely discrete case due to the form of equations~(\ref{pdexpsyst}).
\end{remark}

\section{Acknowledgements}

The author wishes to thank Prof. Alexander Bobenko and TU Berlin for hospitality. Research is partially funded by the Deutsche Forschungsgemeinschaft (DFG) --
TRR109 Collaborative Research Center ``Discretization in Geometry and Dynamics'' (Project ID: 195170736).


\begin{thebibliography}{99}

\bibitem{ShJa}
A.\,B.\,Shabat, R.\,I.\,Yamilov. Exponential systems of type I and the Cartan matrices. {\it Preprint Ufa: BFAN USSR}, 1981. (In Russian)

\bibitem{Le}
A.\,N.\,Leznov. On the complete integrability of a nonlinear system of partial differential equations in two-dimensional space. {\it Theoretical and Math. Phys.},
{\bf 42} (1980), 225--229.

\bibitem{MOP}
A.\,V.\,Mikhailov, M.\,A.\,Olshanetsky, A.\,M.\,Perelomov. Two-dimensional generalized Toda Lattice. {\it Commun.\,Math.\,Phys.}, {\bf 79} (1981), 473--488.

\bibitem{LSSh}
A.\,N.\,Leznov, V.\,G.\,Smirnov, A.\,B.\,Shabat. The group of internal symmetries and the conditions of integrability of two-dimensional dynamical systems.
{\it Theoretical and Math. Phys.}, {\bf 51} (1982), 322--330.

\bibitem{Hi1}
R.\,Hirota. Discrete analogue of a generalized Toda equation. {\it J.\,Phys.\,Soc.\,Jpn.}, {\bf 50} (1981), 3785--3791.

\bibitem{Hi2}
R.\,Hirota. Discrete two-dimensional Toda molecule equation. {\it J.\,Phys.\,Soc.\,Jpn.}, {\bf 56} (1987), 4285--4288.

\bibitem{Wa}
R.\,S.\,Ward. Discrete Toda field equations. {\it Phys.\,Lett.\,A}, {\bf 199} (1995), 45-48.

\bibitem{Do}
A.\,Doliwa. Geometric discretisation of the Toda system. {\it Phys.\,Lett.\,A}, {\bf 234} (1997), no.~3, 187--192.

\bibitem{AS}
V.\,E.\,Adler, S.\,Ja.\,Startsev. Discrete analogues of the Liouville equation. {\it Theoretical and Math. Phys.}, {\bf 121} (1999), 1484--1495.

\bibitem{IH}
R.\,Inoue, K.\,Hikami. The lattice Toda field theory for simple Lie algebras: Hamiltonian structure and $\tau$-function.
{\it Nucl.\,Phys.\,B}, {\bf 581} (2000), 761--775.

\bibitem{Ha06}
I.\,T.\,Habibullin. $C$-series discrete chains. {\it Theoretical and Math. Phys.}, {\bf 146} (2006), no.~2, 170--182.

\bibitem{KNS}
A.\,Kuniba, T.\,Nakanishi, J.\,Suzuki. $T$-systems and $Y$-systems in integrable systems. {\it J.\,Phys. A: Math. Theor.},
{\bf 44} (2011), 103001.

\bibitem{HZY}
I.\,Habibullin, K.\,Zheltukhin, M.\,Yangubaeva. Cartan matrices and integrable lattice Toda field
equations. {\it J.\,Phys. A: Math. Theor.}, {\bf 44} (2011), 465202.

\bibitem{GHY}
R.\,Garifullin, I.\,Habibullin, M.\,Yangubaeva. Affine and finite Lie algebras and integrable Toda field equations on discrete time-space.
{\it SIGMA}, {\bf 8} (2012), 062.

\bibitem{Sm12}
S.\,V.\,Smirnov. Semidiscrete Toda lattices. {\it Theoretical and Math. Phys.}, {\bf 172} (2012), 1217--1231.

\bibitem{CHHL}
X\,Chang, Y.\,He, X.\,Hu, S.\,Li. Partial-Skew-Orthogonal Polynomials and Related Integrable Lattices with Pfaffian Tau-Functions.
{\it Comm.\,Math.\,Phys.} {\bf 364} (2018), 1069--1119.

\bibitem{YF}
Y.\,Yin, W.\,Fu. Integrable semi-discretisation of the Drinfel'd--Sokolov hierarchies. {\it Nonlinearity} {\bf 35} (2022), 3324-3357.

\bibitem{HZhS}
I.\,T.\,Habibullin, N.\,Zheltukhina, A.\,Sakieva. Discretization of hypebolic type Darboux integrable equations preserving integrability. {\it J.\,Math.\,Phys.},
{\bf 52} (2011), no.~9, 093507.

\bibitem{Sm15}
S.\,V.\,Smirnov. Darboux integrability of discrete two-dimensional Toda lattices. {\it Theor.\,Math.\,Phys.}, {\bf 182} (2015), 189--210.

\bibitem{SS}
V.\,V.\,Sokolov, S.\,Ya.\,Startsev. Symmetries on nonlinear hyperbolic systems of the Toda chain type. {\it Theor.\,Math.\,Phys.},
{\bf 155} (2008), 802--811.

\bibitem{De}
D.\,K.\,Demskoi. Integrals of open two-dimensional lattices. {\it Theor.\,Math.\,Phys.}, {\bf 163} (2010), 466--471.

\bibitem{Ni}
Zh.\,Nie. On Characteristic Integrals of Toda Field Theories. {\it J.\,Nonlin.\,Math.\,Phys.}, {\bf 21}, no.~1 (2014), 120--131.

\bibitem{ZISh}
A.\,V.\,Zhiber, N.\,Kh.\,Ibragimov, A.\,B.\,Shabat. Equations of Liouville type. {\it Sov.\,Math.\,Dokl.}, {\bf 20} (1979), 1183--1187.

\bibitem{MS}
D.\,V.\,Millionshchikov, S.\,V.\,Smirnov. Characteristic algebras and integrable exponential systems. {\it Ufa Math.\,J.} {\bf 13}, 2021, no.~2, 41--69.

\bibitem{DT}
D.\,K.\,Demskoi, D.\,T.\,Tran. Darboux integrability of determinant and equations for principal minors. {\it Nonlinearity},
{\bf 29} (2016), no.~7, 1973--1991.

\bibitem{Go1}
E.\,Goursat. {\it Le\c{c}ons sur l'integration des \'equations aux d\'eriv\'ees partielles du second ordre \`a deux variables ind\'ependantes}, vol.~1, 2. Hermann,
Paris, 1896, 1898.

\bibitem{Ha1}
I.\,T.\,Habibullin. Characteristic algebras of fully discrete hyperbolic type equations. {\it SIGMA}, {\bf 1} (2005), 23, 9 pp.

\bibitem{HP}
I.\,T.\,Habibullin, A.\,Pekcan. Characteristic Lie algebra and classification of semidiscrete models. {\it Theoretical and Math. Phys.}, {\bf 151} (2007),
no.~3, 781--790.

\bibitem{HSZ}
N.\,A.\,Zheltukhina, A.\,U.\,Sakieva, I.\,T.\,Habibullin. Characteristic Lie algebra and Darboux integrable discrete chains.
{\it Ufimskij Matem.\,Zhurn.}, {\bf 2}, 2010, no.~4, 39--51. (In Russian)

\bibitem{HPo}
I.\,Habibullin, M.\,Poptsova. Classification of a subclass of two-dimensional lattices via characteristic Lie rings.
{\it SIGMA}, {\bf 13} (2017), 073.

\bibitem{HKS}
I.\,T.\,Habibullin, M.\,N.\,Kuznetsova, A.\,U.\,Sakieva. Integrability conditions for two-dimensional Toda-like equations. {\it J.\,Phys. A: Math. Theor.},
{\bf 53} (2020), 395203.

\bibitem{HK20}
I.\,T.\,Habibullin, T.\,N.\,Kuznetsova. A classification algorithm for integrable two-dimensional lattices via Lie--Rinehart algebras.
{\it Theoretical and Math. Phys.}, {\bf 203} (2020),
no.~1, 569--581.

\bibitem{HKh}
I.\,T.\,Habibullin, A.\,R.\,Khakimova. Characteristic Lie algebras of integrable differential-difference equations in 3D.
{\it J.\,Phys. A: Math. Theor.}, {\bf 54} (2021), 295202.

\bibitem{HK21}
I.\,T.\,Habibullin, M.\,N.\,Kuznetsova. An algebraic criterion of the Darboux integrability of differential-difference equations and systems.
{\it J.\,Phys. A: Math. Theor.}, {\bf 54} (2021), 505201.

\bibitem{Pe}
S.\,S.\,Pervykh. Characteristic algebras for two-dimensional Toda lattices. {\it Course paper}, Lomonosov Moscow State University,
Dept. of Mathematics and Mechanics, 2017, unpublished.

\end{thebibliography}
\end{document}